# Electronic structure of $RE_{1-x}A_xMnO_3$ manganite films investigated by magnetic circular dichroism spectroscopy

Yulia Samoshkina[a*], Dmitriy Petrov[a], Andrei Telegin[b], Yurii Sukhorukov[b], Andrei Kaul[c], Igor Korsakov[c]

[a] *Kirensky Institute of Physics, Federal Research Center KSC SB RAS, 660036, Krasnoyarsk, Russian Federation*
[b] *M.N. Mikheev Institute of Metal Physics UB of RAS, 620108, Ekaterinburg, Russian Federation*
[c] *Moscow state University, Moscow, 119991, Russian Federation*

[*]uliag@iph.krasn.ru

Magnetic circular dichroism (MCD) spectroscopy was used to study the features of the electronic structure of an epitaxial $La_{0.7}Ca_{0.3}MnO_3$ film in the range of 1.2 - 4 eV. The study of the temperature behavior of the MCD spectra made it possible to establish a correlation between the magnetooptical and transport properties of the sample. The data obtained were analyzed in comparison with MCD data for polycrystalline manganite films of various $RE_{1-x}A_xMnO_3$ compositions. The MCD spectra of the films were compared with the spectra of the off-diagonal component of the permittivity tensor calculated from the data of the magneto-optical Kerr effect for films of the same composition. A unified set of ground and excited electronic states characteristic of $RE_{1-x}A_xMnO_3$ manganites in the visible and near infrared ranges has been identified. These results are important for a qualitative theoretical description of the electronic structure of strongly correlated magnetic oxides.

## 1. Introduction

Manganites of the $RE^{3+}_{1-x}A^{2+}_{x}Mn^{3+/4+}O_3$ type (RE is a rare earth element, A is an alkaline earth metal) have a wide range of different properties and, therefore, are still being actively studied in various fields of physics [1-4]. However, the nature of the electronic states in this material has not been precisely established. The nature of the relationship between the electronic and magnetic states determining the macroscopic properties of the manganites is also not clear. At the same time, the clarification of these issues will make it possible to control the functional properties of the manganites.

As a rule, the theoretical study of the electronic structure of a magnetic material is based on its optical or magneto-optical data. However, the optical conductivity spectra of the $RE_{1-x}A_xMnO_3$ manganites in the visible and near infrared regions are characterized by only one broad band with a weakly noticeable fine structure [5-7]. This led to the emergence of a large number of different interpretations and ambiguity in the comparison of optical data with calculations of the electronic structure of the manganites.

The calculation of the electronic structure of the manganite films based on data of the magneto-optical Kerr effect (MOKE) also has its drawbacks. It is well known that MOKE data critically dependent on the surface quality of the samples. Therefore, significant changes in the shape of the MOKE spectrum are observed depending on the thickness of the films [8, 9], the concentration of the A element [10], and the type of epitaxial strain in the material [11]. It should be noted that qualitative data from the MOKE spectra can only be obtained in combination with spectroscopic ellipsometry. These data represent the spectral dependences of the off-diagonal component of the permittivity tensor ($\varepsilon_{xy}$), directly reflecting the intrinsic nature of the material. However, the temperature behavior of the calculated off-diagonal component was not studied, which led to the ambiguity of the spectral analysis of $\varepsilon_{xy}$ in terms of electronic transitions.

For these purposes, magnetic circular dichroism (MCD) spectroscopy is much more informative, since it directly reflects the behavior of $\varepsilon_{xy}$. MCD is a magneto-optical effect observed in transmitted light and is characterizing the medium absorption [12]. The MCD spectroscopy of polycrystalline films of $La_{1-x}Sr_xMnO_3$ at x = 0.3 (LSMO) and $Pr_{1-x}Sr_xMnO_3$ at x = 0.2 and x = 0.4 (PSMO) grown on the Y-oxide stabilized zirconium dioxide (YSZ) substrates was studied by us earlier in [13]. The study showed that the shape of the MCD spectra does not depend on the films thickness and the type of the RE element. An analysis of the temperature behavior of the MCD spectra made it possible to establish a correlation between the MCD data of the films and the type of their conductivity. Thus, a more complete set of excited electronic states in materials in the range of 1.2 - 4 eV was revealed [13].

It should be noted that the effect of epitaxial strain on the electronic structure of manganite film was found in [11] by using MOKE spectroscopy. In particular, it was indicated that a large compressive strain generated by the $LaAlO_3$ (LAO) substrate induces a change in the electronic structure of the LSMO film. In the present work, the $La_{0.7}Ca_{0.3}MnO_3$ (LCMO) film grown on the LAO was studied by the MCD method. This LCMO/LAO composition demonstrated a pronounced metal-insulator transition, colossal magnetoresistance and giant MO phenomena in visible and infrared light [14, 15].
The effect of the epitaxial strain on the electronic structure of the LCMO manganite film has been studied. The data obtained made it possible to establish a unified set of electronic states in the manganites of $RE_{1-x}A_xMnO_3$ type in the visible and near infrared ranges.

## 2. Material and methods

The $La_{0.7}Ca_{0.3}MnO_3$ epitaxial film with a thickness of 50 nm was grown on the LAO (001) single-crystal substrate by chemical deposition from vapors of organometallic compounds (MOCVD) as described in [16]. After deposition, the LCMO/LAO was annealed for an hour in an oxygen atmosphere at a temperature of 800 $^0$C. The crystal structure and phase purity of the samples were examined by the X-ray diffraction (XRD). The XRD data were collected on the Bruker D8 ADVANCE diffractometer using CuKα radiation at 40 kV and 40 mA in a step mode (the step size 0.016$^0$ and 10 s counting time per step) over the range from 20 to 80$^0$. The X-ray beam was controlled by a 0.6 mm fixed slit. The XRD data [15] showed that the sample consists of a highly oriented LCMO (001) phase with a lattice parameter c = 0.3857 (1) nm. The Curie temperature ($T_C$) of the LCMO/LAO sample determined using SQUID magnetometry (Quantum Design MPMS-XL7 EC) is ~ 255 K.

The MCD spectra were measured in the normal geometry using a grating monochromator: both the magnetic field and the light beam were directed normal to the sample plane. The modulation of the polarization state of the light wave from the right-hand to the left-hand circular polarization relatively to the magnetic field direction was used. The MCD signal was measured in the spectral range of 1.2- 3.9 eV as the difference between the optical densities of the right and left polarized waves in the transmitted light. Magnetic field reached up to 16 kOe. The accuracy of measurements was about 10$^{-4}$, and the spectral resolution was 0.002-0.006 eV depending on the wavelength. Temperature dependences were measured in the range of 77-300 K. The MCD measurement technique is described in more detail in [17].

## 3. Results and Discussion

The MCD spectra obtained for the LCMO/LAO film at different temperatures are shown in Fig. 1a. The spectra show two main absorption bands of the same sign: a broad band in the region of 1.7 eV and an intense broad band in the region of 3.2 eV. It is noteworthy that when the temperature decreases, an additional band of the opposite sign appears in the MCD spectrum near 2.4 eV (Fig. 1b). At the same time, the intensity of the main bands increases. The shape and temperature behavior of the MCD spectrum for the LCMO/LAO film are identical to those characteristic for the LSMO/YSZ and PSMO/YSZ (x = 0.4) polycrystalline films of various thicknesses [13]. For comparison, the MCD spectra of these films are shown in Fig. 2. In addition, they are similar in shape and temperature behavior to the spectral dependence of $ε_{xy}$ for an LCMO film deposited on NdGaO$_3$ (NGO) substrate with (110) and LSMO film deposited on SrTiO$_3$ (STO) substrate with (100) [18]. As noted for the LSMO/YSZ and PSMO/YSZ (x = 0.4) films [13], the MCD value near 2.4 eV changes its direction at some temperature $T_S$. With a further decrease in temperature, an additional band near 2.4 eV is clearly visible in the MCD spectra. It was established that the $T_S$ temperature is close to the transition temperature of the samples to the conductive region (metal-insulator transition). In the case of the LCMO/LAO film, a similar relationship is observed (Fig.1b). Measurements of electronic transport have shown that the metal-insulator transition temperature ($T_{MI}$) for this film is 262 K [15]. The MCD data revealed $T_S$ value within 240 K (Fig.1b).

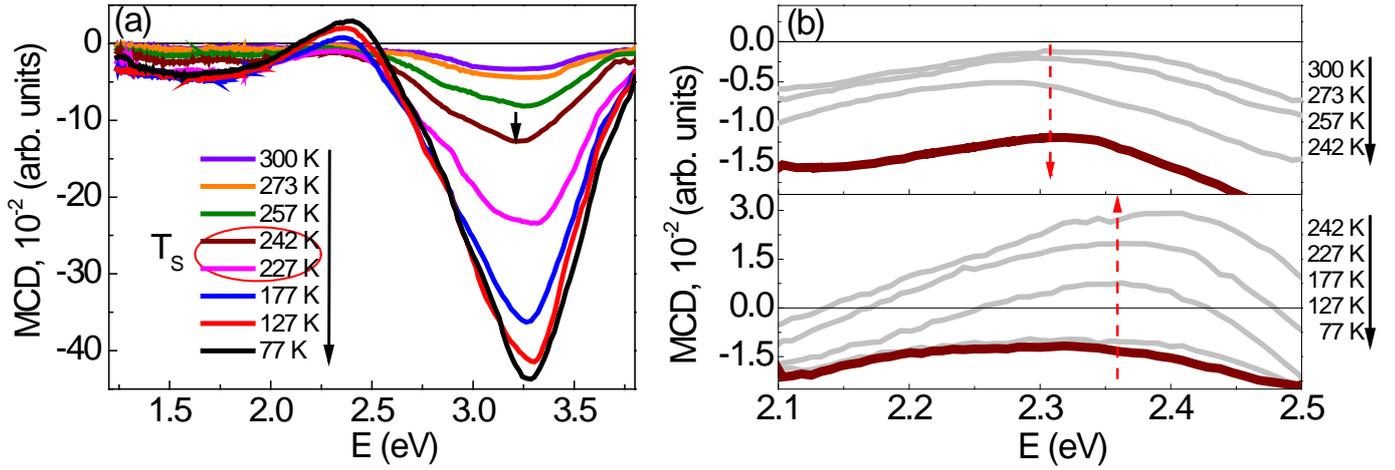

**Fig. 1.** a - Temperature dependencies of MCD spectra of the LCMO/LAO (x = 0.3) film, d = 50 nm; b - The scaled up regions between two main bands in different temperature intervals: upper – T > $T_S$, lower – T < $T_S$. Magnetic field H = 16 kOe.

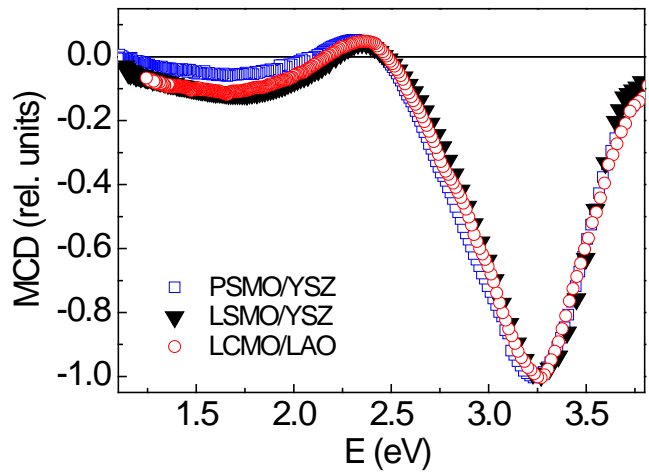

**Fig. 2.** Normalized MCD spectra for the films of LCMO/LAO (x = 0.3), LSMO/YSZ (x = 0.3), and PSMO/YSZ (x = 0.4) at T = 100 K.

As in the case of the LSMO/YSZ and PSMO/YSZ (x = 0.4) polycrystalline films, the MCD spectra of the LCMO/LAO film were decomposed into components and the temperature course of their intensity was traced. The decomposition was performed on the minimal number of the Gaussian-shaped lines. The Gaussian line amplitude (ΔD), position (E), and line width at the half-height (ΔE) were the fitting parameters. Similar to previous results [13], the MCD spectra of the LCMO/LAO sample are well described by four lines in the conducting state and three lines in the insulating state. A typical example is shown in Fig. 3 for the conducting state. The position of all Gaussian lines is presented in Table 1 in comparison with the data of the LSMO/YSZ and PSMO/YSZ (x = 0.4) films taken from [13]. Data are shown in cm$^{-1}$ and in eV for the convenience in case of comparing with data of other authors. Thus, the $E_1$-$E_4$ Gaussian lines represent the $E_1$-$E_4$ MCD bands. The MCD bands position for LCMO/LAO and LSMO/YSZ films coincide within the accuracy. A slight shift in the bands position towards lower energies (beyond the error) is observed for the PSMO/YSZ (x = 0.4) film with a high Sr content. This shift is explained by the screening of the crystal field by the holes density partially localized on the surrounding oxygen ions [19, 20].

The temperature dependences of the $E_1$-$E_4$ bands intensity as well as magnetization data for the LCMO/LAO film are presented in Fig. 4. Despite the fact that MCD signal should reflect the temperature dependence of the sample magnetization, only the $E_4$ band intensity follows the magnetization over the entire temperature range under study. The intensity of the $E_1$ and $E_2$ bands with decreasing temperature follows the magnetization up to T = 155 K, where a kink is observed. Then the intensity of the $E_1$ and $E_2$ bands changes its course. It is noteworthy that the intensity of the $E_3$ band increases with decreasing temperature. However, its temperature course is different from the course of magnetization and $E_4$ band.

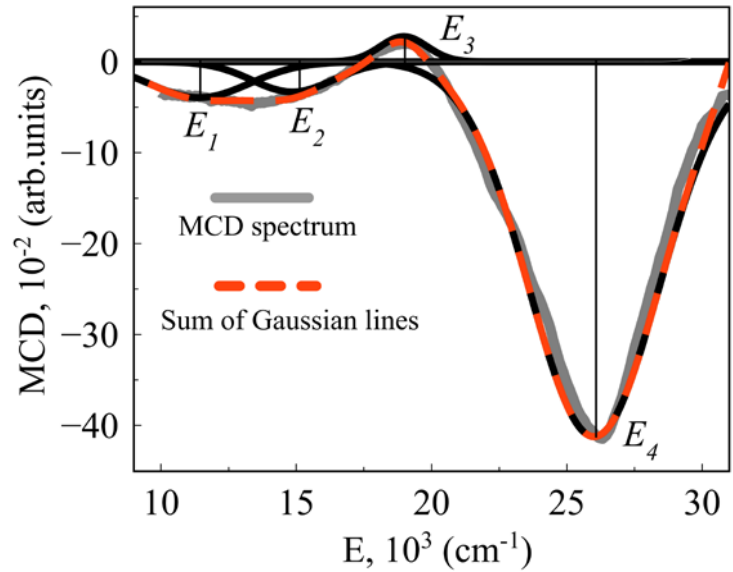

**Fig. 3.** An example of the MCD spectrum decomposition into Gaussian-shaped lines for the LCMO/LAO (x = 0.3) film at T = 100 K.

The temperature behavior of the studied MCD bands coincides with that observed earlier on the polycrystalline films of the LSMO/YSZ and PSMO/YSZ (x = 0.4) of various thicknesses (20 - 150 nm) [13]. Thus, the nature of the observed MCD bands on the LCMO film is similar to that determined for the LSMO/YSZ and PSMO/YSZ (x = 0.4) films (Table 2).

It is known that in the RE$_{1-x}$A$_x$MnO$_3$ manganites, Mn$^{3+}$ and Mn$^{4+}$ ions are located in octahedral complexes of (Mn$^{3+}$O$_6$)$^{9-}$ and (Mn$^{4+}$O$_6$)$^{8-}$, respectively. Considering the effect of only the crystal field in the octahedral complexes (the cluster model), $E_1$, $E_2$, and $E_4$ bands are associated with the spin-allowed d-d transition (Table 2). The energy of these bands is in good agreement with the Tanabe-Sugano diagrams for the $d^3$ ($Mn^{4+}$) and $d^4$ ($Mn^{3+}$) electronic configurations [21].

**Table 1**. The position (E) of the Gaussian lines in the MCD spectra with decomposition at T = 100 K for the LCMO/LAO epitaxial film as well as the LSMO/YSZ and PSMO/YSZ (x = 0.4) polycrystalline films. The error of E position with temperature change is +/- 0.03 eV.

| Composition | $E_1$ | | $E_2$ | | $E_3$ | | $E_4$ | |
| --- | --- | --- | --- | --- | --- | --- | --- | --- |
| | $cm^{-1}$ | eV | $cm^{-1}$ | eV | $cm^{-1}$ | eV | $cm^{-1}$ | eV |
| PSMO/YSZ | 11260 | **1.40** | 14105 | **1.75** | 18880 | **2.34** | 25593 | **3.18** |
| LSMO/YSZ | 11721 | **1.46** | 14850 | **1.84** | 19170 | **2.38** | 25827 | **3.21** |
| LCMO/LAO | 11476 | **1.43** | 14653 | **1.82** | 18983 | **2.36** | 25875 | **3.21** |

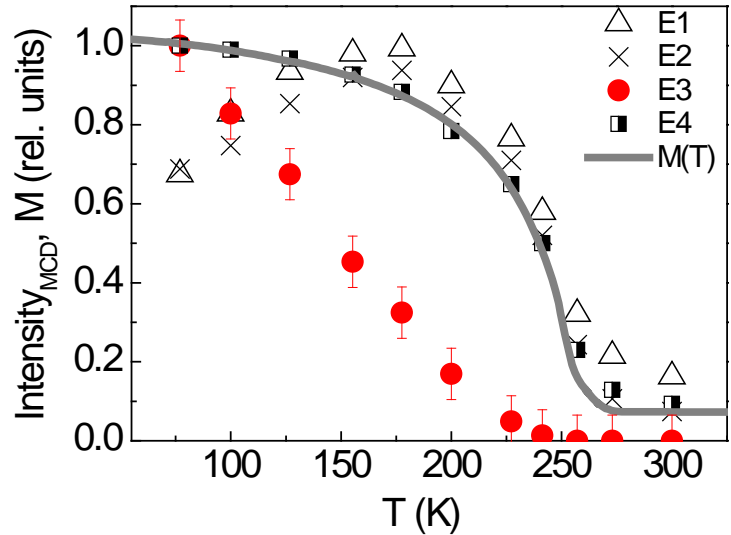

**Fig. 4.** Temperature dependences of the normalized intensity of the MCD bands in comparison with the temperature behavior of the film magnetization (solid curve). Magnetic field H = 16 kOe applied normally to the film plane. The bars show the error in determining the intensity of the bands.

The presence of the kink in the temperature dependence of the $E_1$ and $E_2$ bands intensity (Fig. 4) is explained by the spectral intensity redistribution between excitations of different nature in the region of 0 - 2 eV [13]. This is confirmed by the rearrangement of the optical conductivity and absorption spectra observed in [6, 22, 23] for lanthanum and neodymium manganites in the conducting state. The spectral weight transfer from a higher energy (above 1 eV) to a lower energy (below 1 eV) occurred with decreasing temperature. This phenomenon is not observed in the case of the insulating samples [6, 13], and the intensity of the bands corresponds to the temperature behavior of the magnetization of the sample [13]. In addition, the intensity of bands observed in the case of the insulating samples corresponds to the temperature behavior of the sample magnetization. It should be noted that the reason for this rearrangement has not yet been established. Nevertheless, the data obtained clearly indicate the correlation between the optical (magneto-optical) properties of the LCMO/LAO film in the region of 1 - 2 eV and its conductivity type.

It is noteworthy that the $E_3$ band also correlates with the conductivity type of the film. For the LCMO/LAO film, the $E_3$ band observed only in the conductive region ($T_S < T_{MI}$). Note that this band is absent in the MCD spectra of the PSMO/YSZ (x = 0.2) insulating films [13]. According to band structure calculations of the LSMO/LAO films in [8], this band can be attributed to interband transitions of $Mn^{4+}t_{2g} \rightarrow O_{2p} \rightarrow Mn^{4+}e_g$. The temperature dependence of the intensity of this band is its peculiarity, which has not yet been explained. Thus, a magneto-optical response was found that correlates not with the magnetization, but with the conductivity of the sample. Such behavior is characteristic of dilute oxide semiconductors (for example, based on ZnO [24, 25]). With a completely filled d shell of zinc, the MCD maximum is observed in the region of 3.5 eV for the paramagnetic $Zn_{0.95}Co_{0.05}O$ film. It should be noted that the nature of electronic states in dilute oxide semiconductors is also under discussion.

**Table 2.** Nature of the specified MCD bands and of the possible additional (add.) bands of positive sign in the range of 1 - 5 eV for the LCMO/LAO (x = 0.3) epitaxial film as well as the LSMO/YSZ (x = 0.3) and PSMO/YSZ (x = 0.4) polycrystalline films [13].

| Film | add. band (eV) | $E_1$ (eV) | $E_2$ (eV) | $E_3$ (eV) | $E_4$ (eV) | add. band (eV) |
|---|---|---|---|---|---|---|
| PSMO | < 1 | 1.40 | 1.75 | 2.34 | 3.18 | > 4 |
| LSMO | < 1 | 1.46 | 1.84 | 2.38 | 3.21 | > 4 |
| LCMO | < 1 | 1.43 | 1.82 | 2.36 | 3.21 | > 4 |
| Nature | $Mn^{3+}e_{g1} \rightarrow O_{2p} \rightarrow Mn^{3+}e_{g2}$ interband transitions | $^5E_g \rightarrow {}^5T_{2g}$ d–d transitions in $Mn^{3+}$ ions | $^4A_{2g} \rightarrow {}^4T_{2g}$ d–d transitions in $Mn^{4+}$ ions | $Mn^{4+}t_{2g} \rightarrow O_{2p} \rightarrow Mn^{4+}e_g$ interband transitions | $^4A_{2g} \rightarrow {}^4T_{1g}$ d–d transitions in $Mn^{4+}$ ions | $Mn^{3+}e_g \rightarrow O_{2p} \rightarrow Mn^{4+}e_g$ charge transfer transitions |

Despite the ambiguity of the data obtained using the MOKE spectroscopy, it is worth noting that a band similar to the $E_3$ band near 2.5 eV was observed earlier both in the MOKE spectrum and in the spectrum of $\varepsilon_{xy}$ for the LSMO films deposited on various substrates [8, 9, 11]. Although, the $E_3$ band at T = 40 K is absent in the MOKE spectrum of the studied film [15].

Phenomenologically, the MCD signal repeats the spectral dependence of the real part of $\varepsilon_{xy}$, Re $\varepsilon_{xy}$ [26]. The MCD effect ($\theta_{MCD}$) is defined from the complex value of the Faraday Effect ($\alpha_F$) as

$$\tilde{\alpha}_F = \alpha_F + i\theta_{MCD} = b(\Delta n - i\Delta k) = b\underbrace{\frac{n\varepsilon'_{xy} + k\varepsilon''_{xy}}{n^2 + k^2}}_{\alpha_F} - ib\underbrace{\frac{k\varepsilon'_{xy} - n\varepsilon''_{xy}}{n^2 + k^2}}_{\theta_{MCD}}, \qquad (1)$$

where n is refractive indices, k is absorption indices, b is some coefficient, $\varepsilon'_{xy}$ and $\varepsilon''_{xy}$ are Re $\varepsilon_{xy}$ and imaginary (Im $\varepsilon_{xy}$) part of $\varepsilon_{xy}$, respectively. It is clearly seen that $\theta_{MCD} \sim$ Re $\varepsilon_{xy}$ under the conditions n

<< k. The spectral dependence of Im$\varepsilon_{xy}$, can conventionally be considered a derivative of Re$\varepsilon_{xy}$. It was noticed that the Re$\varepsilon_{xy}$ spectra of the LSMO film exhibit two more bands of the same sign as the $E_3$ band [8, 9, 27]. One band is in the range of 1 - 1.5 eV and the second one is above 4 eV. Thus, two positive bands can be expected in the MCD spectra of the LCMO film at energies below the $E_1$ band and above the $E_4$ band (Table 2) associated with interband transitions of $Mn^{3+}e_{g1} \rightarrow O_{2p} \rightarrow Mn^{3+}e_{g2}$ and charge transfer transitions of $Mn^{3+}e_g \rightarrow O_{2p} \rightarrow Mn^{4+}e_g$, respectively. These conclusions were formed on the basis of theoretical calculations of the electronic band structure of the LSMO and LCMO compounds [5, 8, 28, 29].

It should be noted that in some scientific reports the shape of the Re$\varepsilon_{xy}$ and Im$\varepsilon_{xy}$ spectra for the manganites is confused. This complicates the analysis of the electronic structure of the manganites. Therefore, based on the data obtained, a scheme of the shape of the spectral dependence of Re$\varepsilon_{xy}$ and Im$\varepsilon_{xy}$ for conducting manganites of the $RE_{1-x}A_xMnO_3$ type was compiled (Fig. 5).

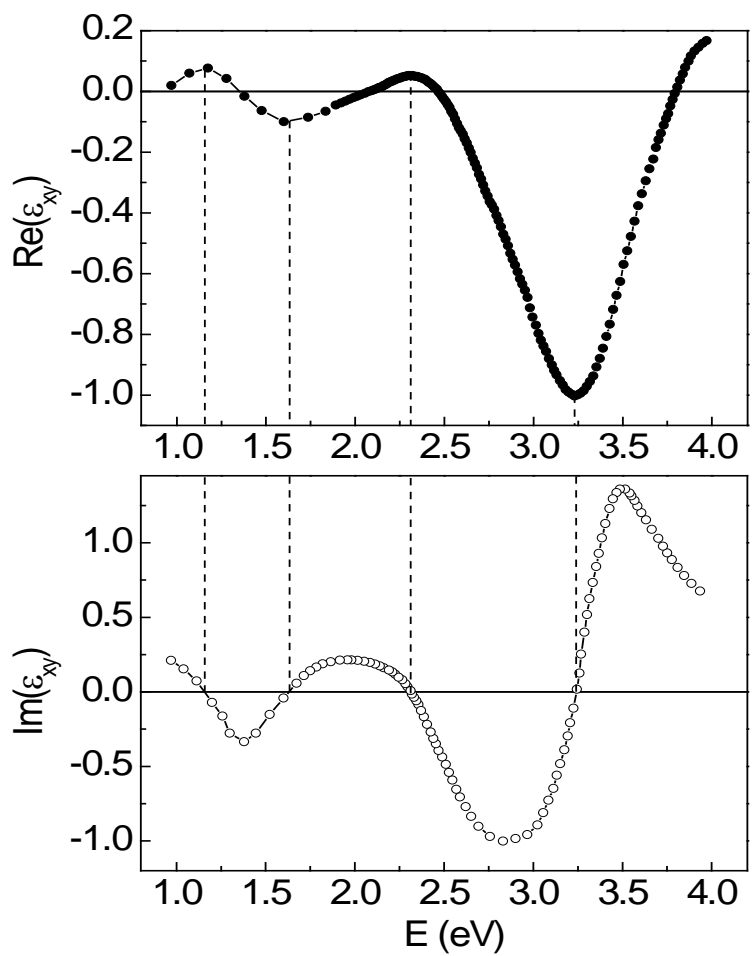

**Fig. 5.** A scheme of the shape of the spectral dependence of Re$\varepsilon_{xy}$ and Im$\varepsilon_{xy}$ for conducting manganites of the $RE_{1-x}A_xMnO_3$ type.

In general, the spectral dependence of the MCD signal of the LCMO/LAO epitaxial film repeats the shape of the MCD spectra of the polycrystalline manganite films of the different composition. Therefore, epitaxial strain does not affect the electronic structure of the manganite films. Moreover, the shape of the MCD spectra of the manganite films coincides with the spectral dependence of MOKE signal observed for the single crystals of $La_{0.85}Ba_{0.15}MnO_3$ and LSMO (x = 0.15-0.25) [30, 31]. This indicates that the revealed features of the electronic structure of the studied films are of general nature and can be extended to any composition of the $RE_{1-x}A_xMnO_3$ manganites of any dimension.

## 5. Conclusions

In summary, it was shown that the shape of the MCD spectra for the $RE_{1-x}A_xMnO_3$ films with the same electrical conductivity type does not depend on the film thickness and the presence or absence of epitaxial strain in the sample. For the films in the insulating state, the MCD spectra contain three absorption bands associated with spin-allowed d-d electron transitions in $Mn^{3+}$ and $Mn^{4+}$ ions. The intensity of these bands corresponds to the temperature behavior of the sample magnetization. An additional band near 2.4 eV appears in the MCD spectra when the films are in the conducting state. With an appropriate instrument sensitivity, another additional MCD band of the same sign can be observed below 1 eV. This band is attributed to interband transitions of $Mn^{3+}e_{g1} \rightarrow O_{2p} \rightarrow Mn^{3+}e_{g2}$. Under these conditions, only the intensity of the high-energy band near 3.2 eV fully corresponds to the temperature behavior of the sample magnetization. It was also found that both types of the films conductivity are characterized by electronic transitions with charge transfer of $Mn^{3+}e_g \rightarrow O_{2p} \rightarrow Mn^{4+}e_g$ localized above 4 eV.

The data obtained indicate a unified set of ground and excited electronic states observed in the visible and near infrared ranges for manganites of the $RE_{1-x}A_xMnO_3$ type. These results are important for a qualitative theoretical description of the electronic structure of the manganites. The revealed correlation effects, in turn, can be used to develop new methods for controlling magnetic and electronic properties of the manganites in general.

**Acknowledgements**

This work was supported by the Russian Science Foundation [grant number 21-72-00061].
The authors are grateful to D.S. Neznakhin and E.A. Stepanova (Ural Federal University, Yekaterinburg, Russian Federation) for the magnetic measurements of the sample.